\documentclass[preprint,prb,showpacs,preprintnumbers,amsmath,amssymb]{revtex4}
\usepackage{graphicx} 
\begin{document}
\title{Quantum Mechanical Simulation of Electronic Transport in Nanostructured Devices by Efficient Self-consistent Pseudopotential Calculation}
\author{\surname{Xiang-Wei} Jiang}
\author{\surname{Shu-Shen} Li}
\author{\surname{Jian-Bai} Xia}
\affiliation{State Key Laboratory for Superlattices and
Microstructures, Institute of Semiconductors, Chinese Academy of
Sciences, P.O. Box 912, Beijing 100083, China}
\author{\surname{Lin-Wang} Wang}
\affiliation{Computational Research Division, Lawrence Berkeley
National Laboratory, Berkeley, California 94720}

\begin{abstract}
We present a new empirical pseudopotential (EPM) calculation
approach to simulate the million atom nanostructured semiconductor
devices under potential bias using the periodic boundary conditions.
To treat the non-equilibrium condition, instead of directly
calculating the scattering states from the source and drain, we
calculate the stationary states by the linear combination of bulk
band method and then decompose the stationary wave function into
source and drain injecting scattering states according to an
approximated top of the barrier splitting (TBS) scheme based on
physical insight of ballistic and tunneling transport. The
decomposed electronic scattering states are then occupied according
to the source/drain Fermi-Levels to yield the occupied electron
density which is then used to solve the potential, forming a
self-consistent loop. The TBS is tested in an one-dimensional
effective mass model by comparing with the direct scattering state
calculation results. It is also tested in a three-dimensional 22 nm
double gate ultra-thin-body field-effect transistor study, by
comparing the TBS-EPM result with the non-equilibrium Green's
function tight-binding result. We expected the TBS scheme will work
whnever the potential in the barrier region is smoother than the
wave function oscillations and if it does not have local minimum,
thus there is no multiple scattering as in a resonant tunneling
diode, and when a three-dimensional problem can be represented as a
quasi-one-dimensional problem, e.g., in a variable separation
approximation. Using our approach, a million atom non-equilibrium
nanostructure device can be simulated with EPM on a single processor
computer.
\end{abstract}

\date{\today}
\maketitle

\section{Introduction}

According to the roadmap of the Semiconductor Industry Association
\cite{ITRS}, MOSFET (metal oxide semiconductor field-effect
transistor) channel length will scale down to 22 nm by 2012. In such
nanosized devices, quantum mechanical effects play a big role in
determining the properties of the system. New quantum mechanical
features, like the fact that the electron mean free path is larger
than the device dimensions and the single quantum state levels, can
be used to enhance device performance and form new functionalities
\cite{Ieong04,Taur02,Hiramoto06}. On the other hand, as the size
reduces, new obstacles emerge \cite{Taur02,Iwai04}, such as the
short channel effects, source/drain off-state quantum tunnelling
current, barrier current leakage  and single dopant random
fluctuation \cite{Roy05}. Over the past 20 years, many methods have
been developed to incorporate the quantum mechanical effects into
the device simulation \cite{Shin07,Anantram07,Sverdlov08}. The first
of that is the inclusion of some quantum mechanical effective
potentials in the drift equation based on the gradients of the
charge density \cite{Iafrate81,Ferry93}. Such gradient terms make
the charge density smooth near the Si/SiO2 interface, hence
incorporating some of the quantum mechanical effects. The so called
quantum Poisson drift equation, or quantum hydrodynamic model have
been extensively used for device simulations. However, as the device
size shrunk further, it was realized that the quantum mechanical
wave functions need to be calculated explicitly.

There are many ways to do the Poisson-Schr\"{o}dinger's equation
depending on the problem to be studied and the computational costs.
One way is to calculate the quantum mechanical local density of
states, then apply Boltzmann transport equation based on such
density of states \cite{Jin08, Vasileska06}. However, in the
ballistic size region, the use of Boltzmann equation itself is
questionable. In such cases, the direct solution of the open
boundary condition scattering states based on the Schrodinger's
equation is necessary. For example, this has been done for the 1D
cases using the nonequilibrium Green's function approach base on
tight-binding model \cite{Lake97}. The use of Green's function also
provides a way to incorporate the inelastic scattering processes in
the formalism. Recently, three dimensional devices models with
hundreds of thousands of atoms have been simulated using the
tight-binding model based on the calculation of scattering states
\cite{Luisier06,Luisier09}. But thousands of computer processors are
needed for such direct 3D simulations. There are also effective mass
calculation for 2D systems using the scattering state approach
\cite{Lent90}. It involves the solutions of linear equations in the
dimension of the number of real space grid points. Overall, the
direct simulation for the 3D device model based on quantum
mechanical transport equation remains to be nontrivial. Thus, it
will be very useful if there is a faster way to do the simulations.
One possible approach is to use the stationary eigen states of a
closed system (e.g., periodic system) Schrodinger's equation to
represent the quantum mechanical effects of the scattering states
\cite{Curatola04,Trellakis97,Abramo00,Hanajiri04,de Falco05}. The
calculation of the eigen states for a closed boundary condition
(e.g., periodic boundary condition) problem is much faster than the
calculation of an open boundary condition scattering states. This
approach is plausible in the sense, at the zero bias potential
between the source and drain, the open boundary condition problem is
the same as the closed boundary condition problem. Thus, the closed
boundary condition solution is a good starting point for the open
boundary condition problem. Many of the quantum mechanical effects
have already been represented at this close boundary condition
level. The challenge however is to find a good approximation to get
the charge density from the stationary wave functions when there is
a large bias potential, and to find the corresponding electron
current. The possibility of such an approximation relies on the fact
that the potential profiles for many semiconductor electronic
devices are often relatively simple and smooth, when there is no
local minimum in the potential, thus no pseudo localized states, the
perspective of the coherent multiple scattering is small. Thus, we
will exclude ourselves from cases like the resonant tunneling diode
(RTD), or complicated molecular electronics. The approximation might
be feasible especially in the regime of ballistic transport, here we
define it not only as elastic transport (ignoring electron-phonon
scattering), but also as a current overcomes a barrier, then flushes
through down hill without multiple scattering
\cite{Jin08,Vasileska06}. In the down hill flushing regime, simple
approximations like the WGK approximation can be applied. Note that,
the scattering states satisfy the same Schrodinger's equation as the
stationary states, albeit their boundary conditions are different.
Thus the eigen states might contain the information as needed for a
transport problem (e.g., the local density of states). In this work,
we will present a new way to obtain the scattering states from the
stationary eigen states. As can be seen in the following sections,
our approach is physically intuitive, easy to implement, and tested
to be accurate for the smooth barrier potential cases as described
above.

Besides the open boundary condition and close boundary condition
problem, there is an issue of what Hamiltonian to be used to
describe the electronic property of the system. We will use
empirical pseudopotential method (EPM) as our Hamiltonian to study
the problem. We have mentioned the tight-binding model
\cite{Lake97,Luisier09,Dicarlo03,Bescond04} and effective mass model
\cite{Lent90,Shin07,Anantram07} above. Another often used model is
the the $k\cdot p$  model which can be used to describe the multiple
valence band states \cite{Curatola04,Bescond04,Wang04}. However, as
shown by Esseni and Palestri et al. \cite{Esseni05}, the $k\cdot p$
can significantly misrepresent the electron density of states (DOS)
for a Si inversion layer, and the indirect band gap nature of bulk
Si presents real challenges for the $k\cdot p$/effective-mass-like
models \cite{Dicarlo03}. It has been shown that in some cases, the
use of the full band structures is important in simulating Si
nanodevices \cite{Neophytou08,Sacconi03}. The EPM is an accurate
method to describe the full semiconductor band structures and
electron wave functions. Within EPM, the total electron potential
$V(r)$ is described as a superposition of the spherical atomic EPM
potentials $v_{\alpha}(|r|)$ as $V(r)=\sum_R v_{\alpha}(|r-R|)$,
while $v_{\alpha}(|r|)$ for atom type $\alpha$ is fitted to
experimental bulk band structures, and $R$ is the atomic positions.
The wave functions in the EPM approach are expanded using plane wave
basis sets. The atomic feature of EPM is important for simulating
small nanodevices where the single atomic characters become
significant. EPM can also describe other effects like strain,
heterostructure, and semiconductor alloying and components, which
are the current research topics for Ge/Si and InAs devices. Another
reason to use EPM is one particularly fast algorithm to calculate
the electron eigen states in a periodic boundary condition
situation. This is the Linear combination of bulk band (LCBB) method
\cite{Wang99}. The LCBB uses the bulk band states instead of the
original plane waves as the basis set to expand the electron wave
functions. As a result, the number of the basis function can be
truncated to be less than 10,000 by selecting a finite number of
k-points and band index. The resulting Hamitlonian matrix can be
diagonalized by a single processor computer within an hour. The LCBB
eigen energy is within ~10 meV of the directly calculated results
using the plane wave basis set. The LCBB is an atomistic method
since each individual atom can be replaced by another atom, and it
can describe the strain effects \cite{Wang99}. The computational
time of LCBB method is roughly independent of the system size (since
it depends mostly on the number of Bloch basis functions which in
many case is independent of the system size), thus can be used to
calculate million atom nanostructures \cite{Williamson00}.

\section{Simulation Approach}

The simulation approach follows the typical iterative scheme which
solves the Schr\"{o}dinger equation (1) and Poisson equation (2)
self-consistently.

\begin{equation}
{(-\frac{1}{2}\nabla^2+\emph{V}(\textbf{\emph{r}})+\emph{V}_{str}(\textbf{\emph{r}})+\phi(\textbf{\emph{r}}))\psi_{i}(\textbf{\emph{r}})}=
\emph{E}_{i}\psi_{i}(\textbf{\emph{r}})
\end{equation}

\begin{equation}
{\nabla[\varepsilon(\textbf{\emph{r}})\nabla\phi(\textbf{\emph{r}})]}=
-4\pi[\emph{p}(\textbf{\emph{r}})-\emph{n}(\textbf{\emph{r}})+\emph{N}_{d}^+(\textbf{\emph{r}})-\emph{N}_{a}^-(\textbf{\emph{r}})],
\end{equation}

here $V(r)=\sum_R v_{\alpha}(r-R)$ is the total empirical
pseudopotential of silicon crystal \cite{Wang94}. $V_{str}(r)$ is
the quantum confinement potential representing the SiO$_2$ barrier
layer and the potential well between the source and drain for the
artificial periodic boundary condition. $\phi(r)$ is the
electrostatic potential which is solved by the Poisson equation (2).
$\epsilon(r)$ is the position dependent dielectric constant;
$N_d^+(r)$ and $N_a^-(r)$ are donor and acceptor nuclei charges, to
be treated as continuous charge densities in the current
calculation; $p(r)$ is the hole charge density while $n(r)$ is the
occupied electron charge density. A more detail description of the
device setup and the way to solve the Poisson equation have been
described in Ref.\onlinecite{Jiang09}.

As mentioned before, the main issue to be addressed in the current
paper is to calculate occupied electron density $n(r)$ and the total
current from eigen state pairs \{$E_i,\psi_i(r)$\} of Eq.(1).
Previously \cite{Luo07,Jiang09}, we have used the WKB approximation
as a weight function as well as the local Quasi-Fermi Potential to
solve this problem. In the following subsections we will present a
new approach based on physical intuition, and to be tested
numerically. It is also stable in both ballistic and tunneling
cases. Compared to our previous approaches \cite{Luo07,Jiang09}, the
present approach gives a smoother charge density and
current\cite{Luo07}, and is conceptional in the ballistic regime
instead of the thermal scattering regime \cite{Jiang09}. More
importantly, our numerical tests show that the results are
surprisingly accurate compared to open boundary condition scattering
state calculations. In this paper, our formalism will be presented
in a heuristic style, instead of rigorous derivations from the
original scattering state problem. They are derived based on a few
simple principles and assumptions. They are tested for typical
systems representing the problems we intend to solve. As will be
discussed below, our top of the barrier splitting (TBS) algorithm
will be based on an one-dimensional effective mass algorithm, while
our original eigen-state wave function could be calculated by EMP
(e.g., in subsection B). Thus it represents a hybrid approach. The
effective mass TBS will not devalue the final result to the
effective mass level since the features of the atomistic EPM
calculation, e.g., the atomistic wave function, the non-parabolicity
of the kinetic energy, the multiple valley, will be retained in the
overall procedure.

Our problem can be summarized as  how to use \{$E_i,\psi_i(r)$\} to
occupy the system with a left and a right Fermi energies $E_F^L$ and
$E_F^R$ to get the occupied charge density $n(r)$ and how to
estimate the current $I$. Formally, this problem can be solved by
occupying the scattering states $\psi_S^L(r,E)$,$\psi_S^R(r,E)$ by
the left and right Fermi energies respectively,

\begin{equation}
n(r)=\int
(|\psi_S^L(r,E)|^2f(E-E_F^L)+|\psi_S^R(r,E)|^2f(E-E_F^R))(\frac{\partial
E}{\partial k})^{-1}dE
\end{equation}

\begin{equation}
I=\int (T_s^L(E) f(E-E_F^L)-T_s^R(E) f(E-E_F^R))(\frac{\partial
E}{\partial k})^{-1}dE
\end{equation}

where $f(x)$ is the Fermi-Dirac distribution function for a given
temperature T, and $T_s^L(E)$, $T_s^R(E)$ are the left and right
scattering states transmission coefficients.  The scattering states
are wave functions satisfying the same Schrodinger's equation as in
Eq.(1), but with an open boundary condition \cite{Lekue09}. So, our
question is whether we can use \{$E_i,\psi_i(r)$\} to mimic the
effects of the scattering states $\psi_S^L(r,E)$ and
$\psi_S^R(r,E)$, or at least their energy integrated properties
$n(r)$ and $I$.

Subsection A will describe the one-dimensional splitting algorithm
while subsection B will extend it to three-dimensional case which
will then be incorporated with the LCBB calculation for our device
simulation.

\subsection{One-Dimensional Model}

Suppose that the electron is running along the x direction, here we will base our
formalism on an 1D effective mass Hamiltonian:

\begin{equation}
(-{1\over2 m_x^*} {d\over d x^2}+V(x)) \psi_i(x)=E_i \psi_i(x)
\end{equation}

We will assume $V(x)$ is a smooth function as shown in Fig.1, and it
has a barrier at the center just like the situation in a transistor,
more specifically, the variation is slower than the wave function
oscillations, and there is no local potential minimum in the barrier
region. Now, for each eigen wave function $\psi_i(x)$ under a closed
boundary condition (in our case, a periodic boundary condition), we
will like to break it into the left injecting $\psi_i^L(x)$ and the
right injecting $\psi_i^R(r)$ scattering states. We first require
the scattering states satisfying the charge conservation rule:

\begin{equation}
|\psi_i(x)|^2=|\psi_i^L(x)|^2+|\psi_i^R(x)|^2
\end{equation}

The above requirement Eq.(6) is needed so that at equilibrium
($E_F^L=E_F^R$), occupying $|\psi_i^L(x)|^2$ and $|\psi_i^R(x)|^2$
is the same as occupying $|\psi_i(x)|^2$ as in the original closed
system problem. The occupied electron density as well as the current
density will be obtained from equation (3) and (4).

Now for a 1D potential $V(x)$ shown in Fig.1, we will have a unique
maximum barrier height $V_m$ at $x_m$ (since there is no potential
minimum), and left source potential $V_L$ and right drain potential
$V_R$ as shown in Fig.1. In the following we will distinguish three
cases: ballistic case with $E_i\geqslant V_m$, tunneling case with
$V_L< E_i<V_m$, and stationary case of $V_R < E_i < V_L$.

For the ballistic case, we require that $\psi_i^L(x)$ and
$\psi_i^R(x)$ are the same at $x_m$, i.e.,

\begin{equation}
|\psi_i^L(x_m)|^2=|\psi_i^R(x_m)|^2=\frac{1}{2}|\psi_i(x_m)|^2
\end{equation}

This requirement is necessary, as we will see below, coupled with
the current model, this will guarantee the current of the left and
right scattering states equal. Such equality is needed so at
equilibrium when both $|\psi_i^L(x)|^2$ and $|\psi_i^R(x)|^2$ are
occupied, there is no net current.

We define a transport velocity as
$\upsilon_i(x)=\sqrt{\frac{2|E_i-V(x)|}{m_x^*}}$. Now for the left
injecting wave $|\psi_i^L(x)|^2$, the current becomes ballistic
(flushing down hill) for $x>x_m$. Now we will use a ballistic
approximation, where the current equals
$\upsilon_i(x)|\psi_i^L(x)|^2$. Since the current must be a
constant, independent of x, it thus must be equal to
$\upsilon_i(x_m)|\psi_i^L(x_m)|^2$. Similarly, for the right
injecting wave $|\psi^R_i(x)|^2$, the current becomes
$\upsilon_i(x)|\psi^R_i(x)|^2=\upsilon_i(x_m)|\psi^R_i(x_m)|^2$ for
$x<x_m$. This leads to:

\begin{align}
|\psi_i^L(x)|^2 & =\left\{
\begin{array}
[c]{c}%
\frac{\upsilon_i(x_m)}{2\upsilon_i(x)}|\psi_i(x_m)|^2  ~~~~~~~~~~~~~~~~~~~~~~~~~~(x>x_m)\\
|\psi_i(x)|^2-\frac{\upsilon_i(x_m)}{2\upsilon_i(x)}|\psi_i(x_m)|^2
~~~~~~~~~~~~~(x\leqslant x_m)
\end{array}
\right.  \nonumber\\
&
\end{align}

and

\begin{align}
|\psi_i^R(x)|^2 & =\left\{
\begin{array}
[c]{c}%
|\psi_i(x)|^2-\frac{\upsilon_i(x_m)}{2\upsilon_i(x)}|\psi_i(x_m)|^2~~~~~~~~~~~~~(x>x_m)\\
\frac{\upsilon_i(x_m)}{2\upsilon_i(x)}|\psi_i(x_m)|^2
~~~~~~~~~~~~~~~~~~~~~~~~~~~(x\leqslant x_m)
\end{array}
\right.  \nonumber\\
&
\end{align}

Note, the first equation in Eq.(8) and second equation in Eq.(9)
comes from the current conservation law and the ballistic expression
for the current, while the second euqation in Eq.(8) and first
equation in Eq.(9) come from Eq.(6). Now, the currents for the left
and right scattering states equal to:
$J_i^L=J_i^R={1\over2}\upsilon_i(x_m)|\psi_i(x_m)|^2$. Note that,
the first equation in Eq.(8) and second equation in Eq.(9) can also
be derived from WKB approximation.

For the tunneling case, the eigen energy $E_i$ cross the barrier
potential $V(x)$ at $x_L$ and $x_R$, $x_L<x_m<x_R$ as shown in
Fig.1.

Now for the region $x_L\leqslant x \leqslant x_R$, we consider the
barrier interval $(x,x_m')$ (same for $(x_m',x)$), here $x_m'$ is
the minimum position for $|\psi_i(x)|^2$ within the retion
$[x_L,x_R]$ (note that $x_m$ and $x_m'$ might not be the same, and
$x_m'$ depends on the state index i).  We can assume that the ratio
of decay of $|\psi_i^L(x)|^2$ from x to $x_m'$ is the same as the
ratio of decay of $|\psi_i^R(x)|^2$ from $x_m'$ to x. We also assume
that the amplitude splitting equation of Eq.(7) holds at $x_m'$. We
thus have:

\begin{equation}
\frac{|\psi_i^L(x)|^2}{|\psi_i^L(x_m')|^2}=\frac{|\psi_i^R(x_m')|^2}{|\psi_i^R(x)|^2}
\end{equation}

\begin{equation}
|\psi_i^L(x_m')|^2=|\psi_i^R(x_m')|^2=\frac{1}{2}|\psi_i(x_m')|^2
\end{equation}

From the above two equations (10), (11), and Eq.(6),  we can solve
the $\psi_i^{L,R}(x)$ within the region $[x_L,x_R]$ as,

\begin{equation}
|\psi_i^{L,R}(x)|^2=\frac{1}{2}(|\psi_i(x)|^2\pm
\sqrt{|\psi_i(x)|^4-|\psi_i(x_m')|^4})
\end{equation}

where "+" is for the left injecting wave at $x<x_m'$ and the right
injecting wave at $x>x_m'$, while "-" is for the left injecting wave
at $x>x_m'$ and the right injecting wave at $x<x_m'$.

Outside of the barrier (after the left and right injecting wave function
tunnel out of the barrier), it is assumed that the current becomes
ballistic. In that case, the wave function amplitude equals $J/\upsilon_i(x)$, here J is the tunneling
current. All these lead to:

\begin{align}
\left\{
\begin{array}
[c]{c}%
|\psi_i^R(x)|^2=J_i^R/\upsilon_i(x)~~~~~~~~~~~~~~~~~~~~~~~~~~~~~~~~~~(x<x_L)\\
|\psi_i^L(x)|^2=|\psi_i(x)|^2-|\psi_i^R(x)|^2~~~~~~~~~~~~~~~~~~~~~~(x<x_L)
\end{array}
\right.  \nonumber\\
&
\end{align}

and

\begin{align}
\left\{
\begin{array}
[c]{c}%
|\psi_i^L(x)|^2=J_i^L/\upsilon_i(x) ~~~~~~~~~~~~~~~~~~~~~~~~~~~~~~(x>x_R)\\
|\psi_i^R(x)|^2=|\psi_i(x)|^2-|\psi_i^L(x)|^2~~~~~~~~~~~~~~~~~~(x>x_R)
\end{array}
\right.  \nonumber\\
&
\end{align}

Note, the tunneling currents $J_i^R$, $J_i^L$ for the right and left
scattering states should be the same (e.g., when they are both
occupied, there will be no net current). We will use
$J_i^R=J_i^L={1\over2} |\psi_i^R(x_L)| |\psi_i^L(x_R)| \sqrt{2
m_x^*(V(x_m)-E_i)}$. Another possible choice is:
$J_i^R=J_i^L={1\over2} max(|\psi_i^R(x_L)|^2, |\psi_i^L(x_R)|^2)
\sqrt{2 m_x^*(V(x_m)-E_i)}$. Although, these two choices look a bit
arbitrary, in reality, we found no practical difference in our test
between these two choices, we will thus use the former one.

There seems a singularity problem at the classical turning points
$x_L$ and $x_R$ in the first equation of Eq.(13) and Eq.(14).
However, the charge density measure of this singularity is zero.
More explicitly,
$\int_{x_L-\epsilon}^{x_L}|\psi_i^R(x)|^2dx=J_i^R\sqrt{\frac{m_x^*}{2V'(x_L)}}\int_{x_L-\epsilon}^{x_L}\frac{1}{\sqrt{x-x_L}}dx=J_i^R\sqrt{\frac{m_x^*}{2V'(x_L)}}\cdot
2O(\epsilon^{1/2})$. Thus, in practice, we do not find it a problem.
A finite numerical smooth can remove this singularity. Nevertheless,
a small kink can be observed in Fig.3.

For the third case $V_R < E_i < V_L$, it is stationary, thus
$\psi_i^R(x)=\psi_i(x)$, $\psi_i^L(x)=0$, and $J_i^R=J_i^L=0$.

Now the occupied electron density as well as the total current can
be evaluated as

\begin{align}
\left\{
\begin{array}
[c]{c}%
n(x)=\sum_i |\psi_i^L(x)|^2f(E_i-E_F^L)+|\psi_i^R(x)|^2f(E_i-E_F^R)\\
J_{tot}=\sum_i J_i^L f(E_i-E_F^L)-J_i^R
f(E_i-E_F^R)~~~~~~~~~~~~~~~~~
\end{array}
\right.  \nonumber\\
&
\end{align}

Once we have the electron eigen states $\{E_i,\psi_i(x)\}$, the
occupied electron density as well as the current $I$ can be
evaluated from the above model equations. We will call this the top
of barrier splitting (TBS) model.

To test the validity of the above 1D splitting model, we have
calculated the scattering state exactly using the transfer matrix
(TM) method \cite{Jirauschek09}. For 1D and effective mass
Hamiltonian, this is easy to do. There is no numerical stability
problem caused by the multiple band situation. We have chosen a test
potential $V(x)$ with a form as:

\begin{align}
\left\{
\begin{array}
[c]{c}%
V(x)=V_L ~~~~~~~~~~~~~~~~~~~~~~~~~~~~~~~~~~~~~~~~~~~~~~~~~~~~~~~~~~~~~~~~    (x<x_1)   \\
V(x)=-V_g sin(\frac{(x-x_1)\pi}{x_2-x_1})+(V_R-V_L) \frac{x-x_1}{x_2-x_1} + V_L  ~~~~~~~~~~ ( x_1 <x < x_2)  \\
V(x)=V_R~~~~~~~~~~~~~~~~~~~~~~~~~~~~~~~~~~~~~~~~~~~~~~~~~~~~~~~~~~~~~~~~(x>x_2)
\end{array}
\right. \nonumber \\
&
\end{align}

here $x_1=20$ nm, $x_2=45$ nm, $V_L=0$, $V_R=-0.5$ eV, and effective
mass $m_x^*=0.19$ as for silicon. Note the shape of the potential
can be modified by changing $V_g$ and $V_L-V_R$. We have tested
other potential shapes (but with no potential minimum), and find
similar results as shown below in terms of the accuracy of the TBS
results.

In Fig.2(a), the ballistic wave function splitting results are
presented for two periodic boundary condition (by connecting $x=0$
nm potential to $x=65$ nm potential) eigen states with eigen
energies $E_1=0.618$ eV and $E_2=0.620$ eV. With $V_g=-0.8$ eV, the
$V_m$ is at 0.56 eV. Thus, both of these two states belong to the
ballistic situation. Their decomposed $|\psi_i^L(x)|^2$ are shown in
Fig.2(a) (the $|\psi^R_i(x)|^2$ looks similar from the opposite
side). One can see that the constructed individual $|\psi_i^L(x)|^2$
using Eq(6) can be negative. This might sounds alarming. However,
the summation of these two scattering states (with very close eigen
energies) $|\psi_1^L(x)|^2+|\psi_2^L(x)|^2$ are all positive as
shown in Fig.2(b). Furthermore, the summed result resembles closely
to the TM calculated scattering state wave function at
$E=(E_1+E_2)/2$. Thus, as claimed previously, it is the energy
integrated properties which resemble the directly calculated
results, not the individual TBS scattering states. Note that, for
ballistic case wave functions $|\psi_i(x)|^2$ with $E_i > V_m$, the
eigen states always come in pairs with very close eigen energies
(rather like the sin and cos case). There is however and issue for
low temperature occupation. What happens if $|\psi_i^L|^2$ is
occupied while $|\psi_i^R|^2$ is not. Theoretically, this can be
overcome by increasing the size of the source and drain regions, as
a result, the difference between $E_1$ and $E_2$ will diminish. In
practice, at room temperature, we didn't find this is a problem.
Tests for other ballistic case eigen states show similar behavior as
in Fig.2.

In Fig.3, the decomposed $|\psi_i^L(x)|^2$  and $|\psi_i^R(x)|^2$
scattering states for a tunneling case $|\psi_i(x)|^2$ with
$E=0.1814$ eV are shown. As one can see, the decomposed state
resemble closely the directly TM calculated scattering states. At
$x_L$ for $|\psi_i^R(x)|^2$ and $x_R$ for $|\psi_i^L(x)|^2$, there
is a kink to join the wave functions from Eq(12) and Eq(13) as
discussed above. But in practice, since it happens in such a small
amplitude, and the measure of this singularity is zero, it rarely
matters. Other tunneling states show similar behavior. Note that for
tunneling case, there is no pairing for the periodic boundary
condition eigen states $\{E_i,\psi_i(x)\}$.

In Fig.4, the occupied charge densities calculated from Eq(15) are
shown using $E_F^L=0.6$ eV, $E_F^R=0.1$ eV, and $V_g=-0.8V$ at room
temperature. We see that the TBS gives almost an identical occupied
charge density as the TM directly calculated scattering state
results. Note, here we are not doing any selfconsistent calculation
yet, but such charge density is the first step towards the
selfconsistent calculation (as will be done in the 3D calculation
later in the paper). The calculated current for this case is
$4.44\times 10^{-5}$ a.u. for the TBS model, while it is $4.58\times
10^{-5}$ a.u. for the TM directly calculated scattering state model,
differs by only $3\%$.

Now, we change $V_g$ and calculate the current I from left source to
the right drain. This will give us the I-V curve. The results are
shown in Fig.5. We see that the TBS model and the direct scattering
state calculation yield almost the same results over a wide range of
current amplitudes. This shows the accuracy of our TBS model for
device simulations at least for one dimensional system. Note that,
using Fermi-Diract distribution f(x) in Eq.(15), the small current
in the region of $V_g < -0.9$ eV comes mostly from the
over-the-hill-top ballistic current caused by the $\sim exp(-x)$
Boltzmann distribution. This is because as the barrier height
increase, the tunneling current (below the hill top) decreases
faster than the Boltzmann distribution $exp(-x)$ for the
over-the-hill-top states $\{\psi_i(x)$,$E_i\}$ to be occupied. To
test the approximation on the pure tunneling current, we have also
used an artificial occupation function $f_t(x)$ to replace the
Fermi-Diract distribution in Eq(15). $f_t(x)=1$ for $x<0$,
$f_t(x)=cos^2(x\pi/8)$ for $0<x<4$, and $f_t(x)=0$ for $x>4$. Thus,
for large barrier height, there will be no over-the-hill-top
ballistic current, instead all the current will come from the
tunneling. From Fig.5, we see that this pure tunneling current also
agree well between the TBS result and the direct scattering state
calculation results.

\subsection{Three-Dimensional Model}

In above, we have obtained an excellent 1D algorithm. To extend that
to three-dimensional case, we will first still base on an effective
mass Hamiltonian, and employ the variable separation approximation
between the x direction and y, z directions. Such separation is a
good approximation in cases where there is a fast variation of the
potential $V(x,y,z)$ in the y, z directions, but a relatively slow
variation in the x direction. This is applicable to many device
systems with a narrow current channel. Such approximations have
often been used to solve the 3D Hamiltonian eigen state problem.
Here, we will use it to construct our splitting algorithm. Under the
variable separation approximation, the three-dimensional wave
function can be written as $\psi(r)=\zeta(x)\theta(x,y,z)$ and
$\theta(r)$ satisfies $\int |\theta(r)|^2dydz=1$.  We first require

\begin{equation}
(-\frac{1}{2m_y^*}\frac{\partial^2}{\partial
y^2}-\frac{1}{2m_z^*}\frac{\partial^2}{\partial
z^2}+V(x,y,z))\theta(x,y,z)=E(x)\theta(x,y,z)
\end{equation}

This is a 2D eigen value problem with x as a parameter.  Now, if we
assume $\theta(x,y,z)$ varies slowly with x, thus its first and
second order derivatives in x direction can be ignored, then
satisfying the 3D Schrodinger's equation:

\begin{equation}
(-\frac{1}{2m_x^*}\frac{\partial^2}{\partial
x^2}-\frac{1}{2m_y^*}\frac{\partial^2}{\partial
y^2}-\frac{1}{2m_z^*}\frac{\partial^2}{\partial
z^2}+V(x,y,z))\zeta(x)\theta(x,y,z)=\epsilon \zeta(x)\theta(x,y,z)
\end{equation}

is equivalent to satisfying the 1D equation:

\begin{equation}
(-\frac{\hbar^2}{2m_x^*}\frac{\partial^2}{\partial
x^2}+E(x))\zeta(x)=\epsilon \zeta(x)
\end{equation}

Now if by using other eigen state solvers (e.g., LCBB) we have
obtained a three-dimensional eigen state pair $\{\epsilon,\psi(r)\}$
(we have droped the band index i for simplificty), we can then write
$\zeta(x)$  as:

\begin{equation}
\zeta^2(x)=\int|\psi(x,y,z)|^2dydz
\end{equation}

and $\theta(r)$ as $\theta^2(r)=\psi^2(r)/\zeta^2(x)$. Once the
one-dimensional effective potential $E(x)$ is known, we can use the
1D algorithm discussed in the previous section to separate the 1D
$\zeta(x)$ into $\zeta_L(x)$ and $\zeta_R(x)$, which in turn
separate $\psi(r)$ as ($\theta(x,y,z)$ needs not to be separated):

\begin{align}
\left\{
\begin{array}
[c]{c}%
|\psi_L(r)|^2=\frac{|\zeta_L(x)|^2}{|\zeta(x)|^2}|\psi(r)|^2\\
|\psi_R(r)|^2=\frac{|\zeta_R(x)|^2}{|\zeta(x)|^2}|\psi(r)|^2
\end{array}
\right.  \nonumber\\
&
\end{align}

Thus, in practice, if we already have the three dimensional eigen
states $\{\epsilon, \psi\}$, all we need is to get the corresponding
1D effective potential $E(x)$. Note the current in x direction will
be the same as from the 1D formula because $\theta^2(x,y,z)$
normalize to 1 over y and z at any x. We will ignore any current in
the y and z directions. For the effective mass Hamiltonian, $E(x)$
can be obtained from Eq.(17) as:

\begin{equation}
E(x)=\frac{\int\psi^*(r)(-\frac{1}{2m_y^*}\frac{\partial^2}{\partial
y^2}-\frac{1}{2m_z^*}\frac{\partial^2}{\partial
z^2}+V(x,y,z))\psi(r)dydz}{\int|\psi(r)|^2dydz}=E_K^{yz}(x)+E_V(x)
\end{equation}

\subsection{LCBB calculation for the 3D model}

So far, we have only used  effective mass Hamiltonian to derive our
TBS algorithm from $\psi_i(r)$ to $\psi_i^R(r)$ and $\psi_i^L(r)$.
However, we will like to use atomistic Hamiltonian (e.g.,EPM) to
obtain $\psi_i(r)$. One might ask whether the use of effective mass
model derived formula will devalue the atomistic result into
effective mass level. The answer is no. The reason is that the
atomic features are built in $\psi_i(r)$ by the atomistic
Hamiltonian, while the separation algorithm is built on the smooth
part (envelop function part) of $\psi_i(r)$, which can be described
by the effective mass formalism. Using the linear combination of
bulk band method (LCBB method) \cite{Wang99} to solve the electronic
eigen states $\{E_i,\psi_i(r)\}$, the wave function $\psi(r)$ will
be expanded by the Bloch states of the constituent bulk solids:

\begin{equation}
\psi(r)=\sum_{n,k}C_{n,k}u_{n,k}(r)e^{ik\cdot r}
\end{equation}

where the periodic part $u_{n,k}(r)$ of the Bloch function is described by the
plane wave functions as

\begin{equation}
u_{n,k}(r)=\frac{1}{\sqrt{V_0}}\sum_G A_{n,k}(G)e^{iG\cdot r}
\end{equation}

Here, $n$ is the band index, and $k$ is the supercell
reciprocal-lattice vector defined within the first BZ of the silicon
primary cell while $G$ is the reciprocal lattice of the primary cell
chosen within an energy cutoff. The Hamiltonian matrix elements are
evaluated within the basis set $\{u_{n,k}(r)e^{ik\cdot r}\}$, and
the resulting Hamiltonian matrix is diagonalized to yield
$\{C_{n,k}\}$.

We now go back to yield the 1D potential $E(x)$ in Eq(22) using the
LCBB calculations, keep in mind that the separation formalisms in
the subsections A and B correspond to the envelop function
properties with the atomic features (e.g., $u_{n,k}(r)$) of the wave
functions and the potentials  already being removed (or smoothed
out). The one-dimensional integrated potential $E_V(x)$ in Eq.(22)
can be calculated directly from the electrostatic potential
$\phi(r)$ and the quantum confinement potential $V_{str}$ in Eq(1)
as

\begin{equation}
E_V(x)=\frac{\int|\psi(r)|^2(\phi(r)+V_{str}(r))dydz}{\int|\psi(r)|^2dydz}
\end{equation}

The electrostatic potential caused by the carried charge density
$\phi(r)$ is generally smooth. The one-dimensional kinetic energy
$E_K^{yz}(x)$ is the effective mass kinetic energy, which involves
the Laplacian on the envelope function of the wave function (this is
different from the true kinetic anergy of the whole wave function as
in Eq.(1)). In our LCBB representation, envelope function part of
the wave function corresponds to the k component in Eq.(23), not the
G component in Eq.(24). Thus if we define
$\widetilde{\psi}(r)=(-\frac{1}{2m^*_y}\frac{\partial^2}{\partial
y^2}-\frac{1}{2m^*_z}\frac{\partial^2}{\partial
z^2})|_{envelop}\psi(r)$, then under LCBB expansion, we have:

\begin{equation}
\widetilde{\psi}(r)=\sum_{n,k}(\frac{1}{2m_y^*}(k_y-k_0(k))^2+\frac{1}{2m_z^*}(k_z-k_0(k))^2)C_{n,k}u_{n,k}e^{ik\cdot
r}
\end{equation}

here $k_0(k)$ is the bottom of valley point for each k point. For
example, in the Si case studied in the current paper, $k_0$ is the
0.83X point of the bottom of conduction band. Note that evaluating
$\widetilde{\psi}(r)$ adds no computational expense to the LCBB
routine. Once we have $\widetilde{\psi}(r)$, the one-dimensional
kinetic energy $E_K^{yz}(x)$ can be calculated by

\begin{equation}
E_K^{yz}(x)=\frac{\int\psi^*(r)\widetilde{\psi}(r)dydz}{\int|\psi(r)|^2dydz}
\end{equation}

This concludes our full TBS algorithm using LCBB: first to calculate
the eigen states $\{\epsilon_i, \psi_i(r)\}$ under LCBB, then use
Eqs.(25)-(27) to obtain the 1D potential $E(x)$, then use
Eqs.(20),(21) and 1D formalism to separare $\psi_i(r)$ into
$\psi_i^L(r)$ and $\psi_i^R(r)$. The calculation of the carrier
charge density follow naturally from analogues equation of Eq(15),
and the selfconsistent calculation is done via Eq.(2). Finally, the
current is evaluated from Eq.(15).

\section{Simulation for 22-nm double gate ultra-thin-body field-effect transistor}

Following the ITRS 22 nm technology node, we present a
three-dimensional atomistic quantum mechanical simulation on a 22 nm
double gate ultra-thin-body field-effect transistor using the above
presented three-dimensional TBS model. Previously, this kind of
nanodevice has been studied by a TB and NEGF approach
\cite{Luisier08}. Here, to achieve a comparison between our TBS
model and NEGF model, we use the same structural and material
parameters as in Ref. 41. Fig. 6 shows the structure and parameters
of the simulated device. The channel direction is along (100). The
gate length and silicon body thickness is $L_G=22$ nm and
$t_{SI}=4.9$ nm respectively. The thickness of the oxide is
$t_{OX}=1.3$ nm. The channel region is undoped while the source and
drain region is highly $n+$ doped as $N_d=1\times 10^{20} cm^{-3}$.
The drain bias is fixed to be $V_D=1.0V$ in the simulation. The
Schr\"{o}dinger equation (1) is solved in the whole device region
with 0.8 million atoms using the LCBB method. An artificial periodic
boundary condition is used connecting the left and right ends of the
device with an artificial potential barrier between them. The
resulting eigen states $\{E_i,\psi_i(r)\}$ are occupied by the
source/drain Fermi-Levels by the three-dimensional algorithm in the
previous section. Then the occupied electron density is used in the
Poisson equation (2) to form a self-consistent calculation. The
Poisson equation (2) is only solved within the box of the blue
dashed line shown in Fig.6. A typical Pulay DIIS charge mixing
iteration scheme is used to accelerate the speed of the
self-consistent calculation. For more details for how to solve the
Poisson equation and its boundary conditions, please see
Ref.\onlinecite{Jiang09}. In the LCBB calculation, k-points are
chosen from six X-valleys (two X100, two X010 and two X001) and in
total there are about 7000 k-points included in the LCBB expansion.
Two conduction bands are selected in the LCBB basis set. The
calculation is carried out on a computer with a single CPU.

Fig. 7 and 8 shows how the TBS works for such LCBB calculated
million atom 3D system. For ballistic and tunneling case, the black
solid line in Figs. 7 and 8 indicates the total wave of the eigen
state while the red and blue lines with symbols indicate the right
and left running (tunneling) waves. As can be seen from the figure,
the two separated parts coincide very well with the total wave
function at both side of the barrier. For the ballistic case as
shown in Fig. 7, the right running and left running waves are
separated at the top of the barrier and then are injected
ballistically into the other sides of the barrier. For the tunneling
case as shown in Fig. 8, the right and left tunneling waves separate
at the minimum wave function point, then decay exponentially to the
other side of the barrier. Note that unlike in the 1D case shown in
Figs.1 to 3, the potential profile $E(x)$ in the 3D case, hence the
potential maximum point $x_m$ depends on the state index i.
Nevertheless, the separating algorithm remains the same.

Fig. 9 shows the converged occupied electron density and local
conduction band profile (which equals $\phi(r)+E^{bulk}_{CBM}$) for
$V_g=0.0-0.8V$. As can be seen from the figure, in low gate voltage
cases, the electron density decays very fast from the source/drain
to the channel. The electron density at small gate bias decays
$10^7$ order from the source/drain region to the channel region. As
the gate bias gets higher, the barrier in the channel is pushed down
leading to a significantly increased charge density in the channel.
It should be noted that the maximum point of the barrier moves
towards the source side as the gate potential further pushes the
barrier down.

Finally, Fig. 10 compares the I/V curves of our splitting model and
the NEGF incorporated $sp^3d^5s^*$ tight-binding model
\cite{Luisier06,Luisier08}. The I/V data of the TB+NEGF model is
from Ref.41. Despite of the different Hamiltonians used, and
different numerical details in solving the Poisson equations, the
results aggree excellently. Table.I gives a detailed comparison of
some key device performance parameters. As can be seen from the
table, there is just a $12.7\%$ difference between the ON currents
of the two models and only 10 mV difference between the threshold
voltages. Both two models give exactly the same sub-threshold swing
$S=63mV/dec.$ which is defined as $S=\frac{dV_g}{dlog_{10}I_d}$ in
sub-threshold region. This value ($S=63mV/dec.$) is close to the
theoretical limit of $kT\cdot ln10=60meV$, indicating a very
effective gate control in such a double-gate ultra-thin-body device.

Note that the TB+NEGF model solves the transport problem based on an
open boundary condition while our splitting model is based on a
periodic boundary condition and solves the same problem from the
eigen states. From the comparisons, we see that these two methods
lead to almost the same results. However, the computational costs
for these two models are significantly different, while thousands of
computer processors are used for the TB+NEGF approach, a single
processor is used for our TBS approach. Overall, this is an evidence
of the validity of our TBS model for 3D systems.

\section{Summary}

In summary, we have presented a new empirical pseudopotential
calculation approach to simulate million atom nanostructure
semiconductor devices using the periodic boundary conditions. To
treat the non-equilibrium condition, instead of calculating the
scattering states from the source and drain, we calculated the
stationary states by the linear combination of bulk band method and
then separated the whole wave function into source and drain parts
according to a top of the barrier splitting (TBS) scheme based on
the physical insight of ballistic and tunneling transport. The
separated electronic states were then occupied according to the
source/drain Fermi-Levels to yield the occupied electron density
which is then used in the Poisson Equation solver to form a
self-consistent calculation. The validity of TBS was verified by a
comparison between the TBS calculation and scattering states
calculation in a 1D effective mass model. It is also verified by
comparing our LCBB-TBS calculation for a 3D double-gate field-effect
transistor with a non-equilibrium Green's function tight-binding
result. However, instead of using thousands of processors, our
method can be calculated using a single processor. Thus, this method
provides a fast way to simulate the non-equilibrium million atom
nanostructure problems.

\section*{Acknowledgment}
This work was supported by the National Basic Research Program of
China (973 Program) grant No. G2009CB929300 and the National
Natural Science Foundation of China under Grant Nos. 60821061 and
60776061. L.W.W is funded by the U.S. Department of Energy BES,
office of science, under Contract No. DE-AC02-05CH11231.

\newpage

\begin{figure}
\centering
\includegraphics[width=5.4in,height=5.4in,keepaspectratio]{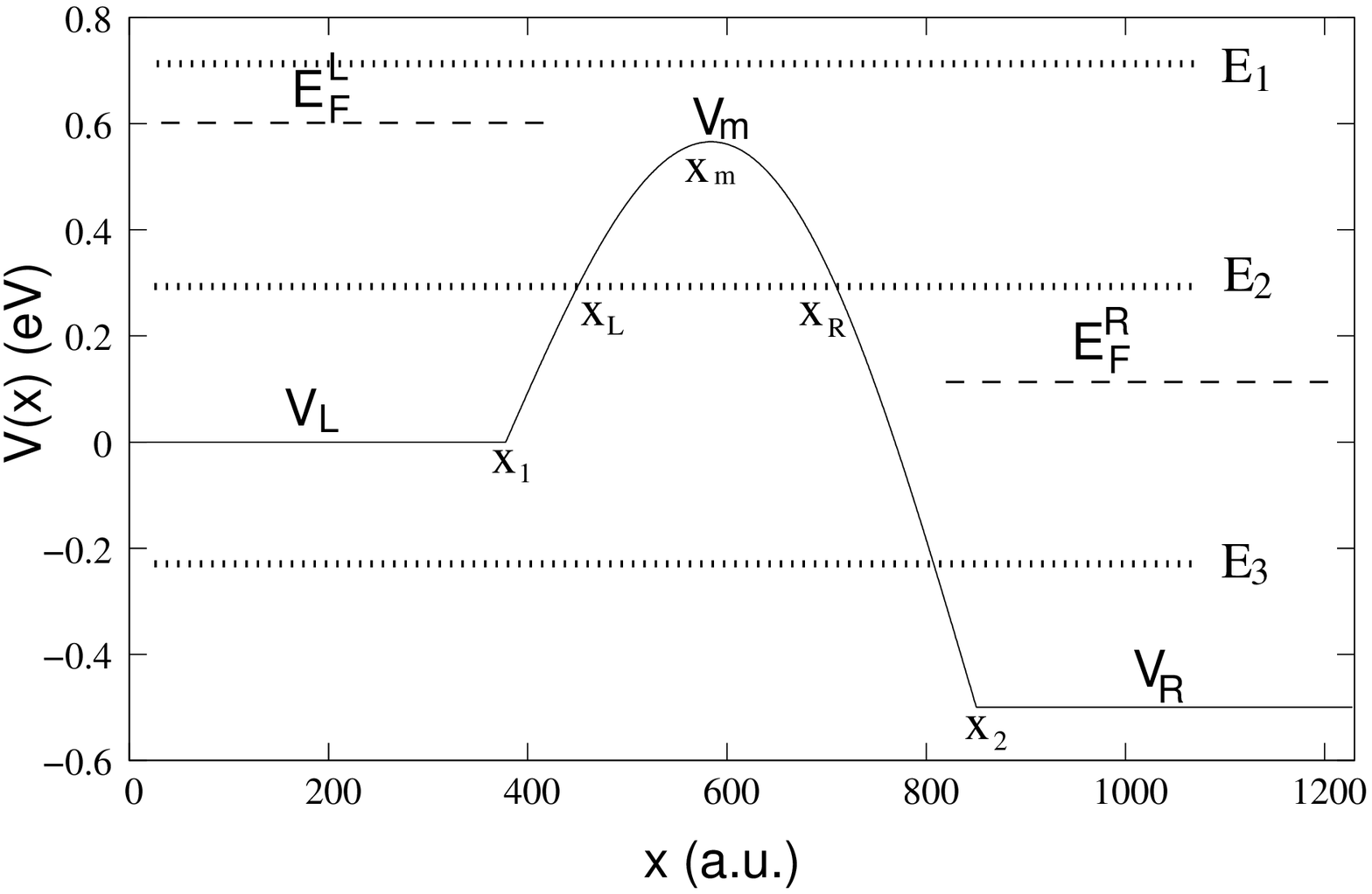}
\caption{ The potential barrier and different energy levels.}
\end{figure}

\begin{figure}
\centering
\includegraphics[width=5.4in,height=5.4in,keepaspectratio]{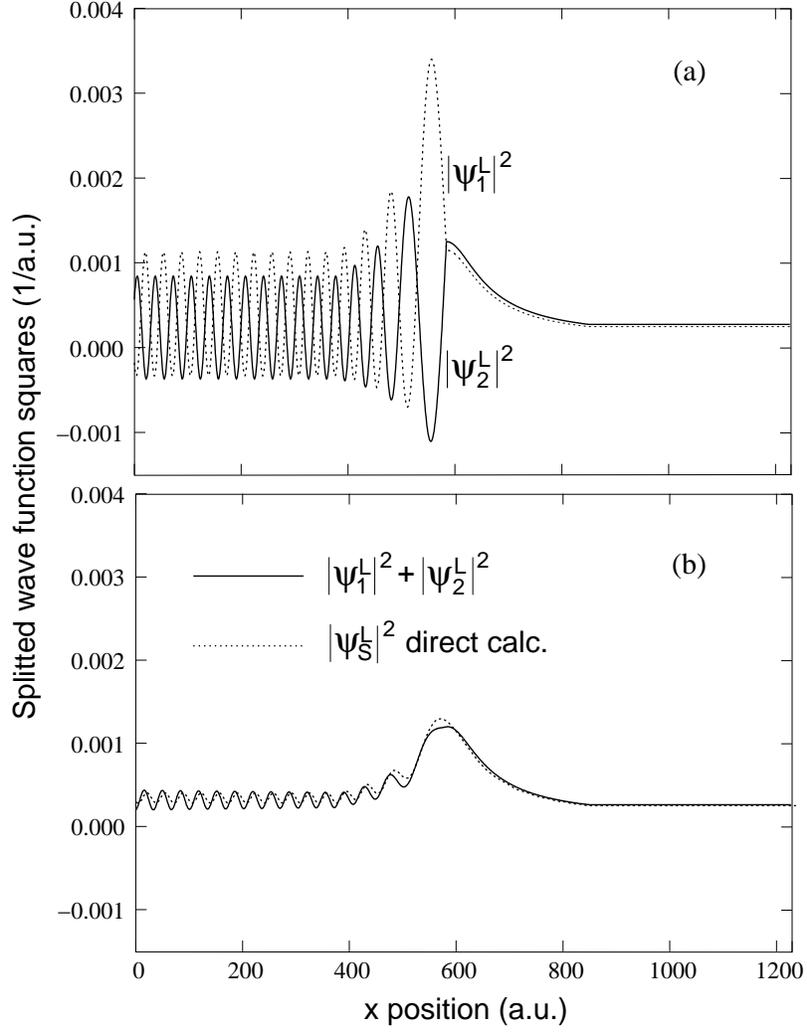}
\caption{ The splitted wave functions for the ballistic case (a) and
compared with the direct calculated scattering states (b).}
\end{figure}

\begin{figure}
\centering
\includegraphics[width=5.4in,height=5.4in,keepaspectratio]{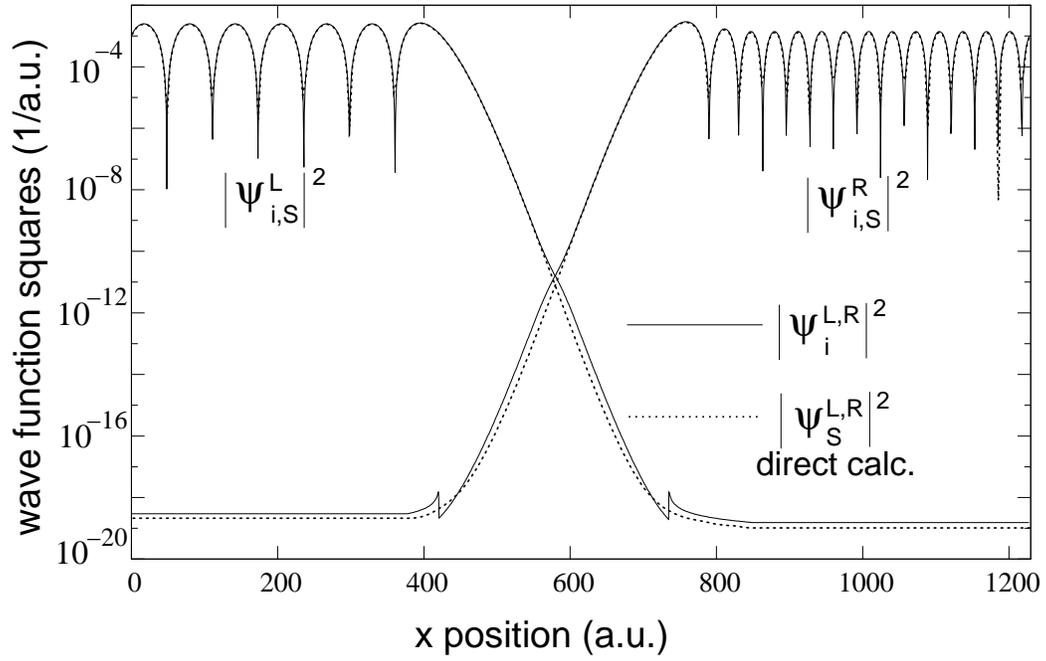}
\caption{ The splitted wave functions for the tunneling case and
compared with the direct calculated scattering states.}
\end{figure}

\begin{figure}
\centering
\includegraphics[width=5.4in,height=5.4in,keepaspectratio]{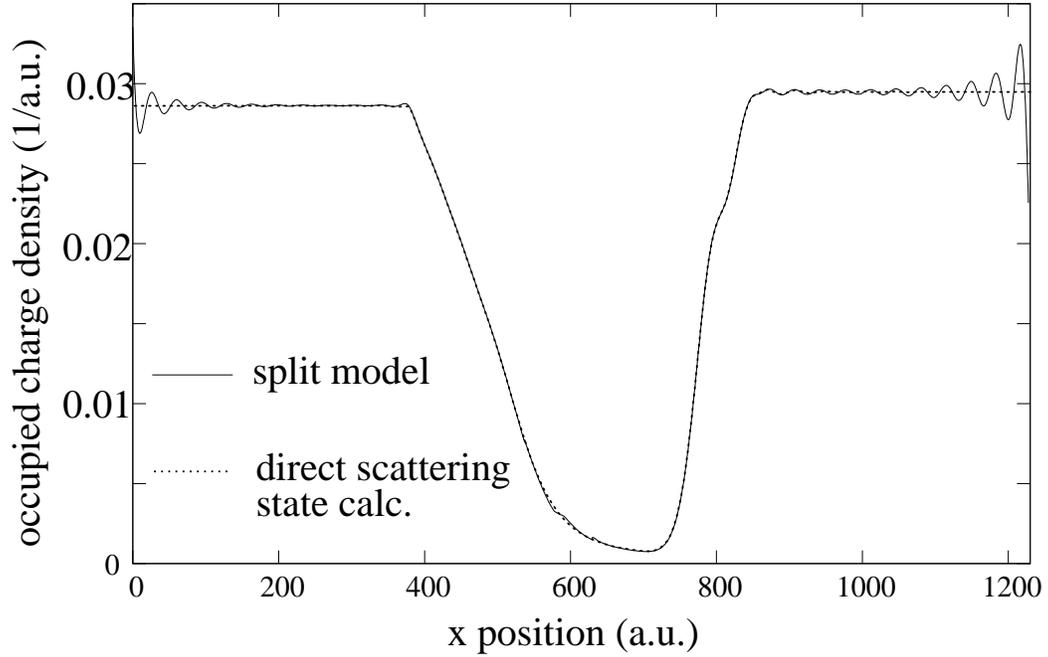}
\caption{ The occupied charge density using equation (14) with the
splitted wave functions in comparison with the charge density
calculated with directly calculated scattering state using equation
(3). The oscillation at the two ends is due to the artificial
periodic boundary condition used to calculate $\psi_i(x)$.}
\end{figure}

\begin{figure}
\centering
\includegraphics[width=5.4in,height=5.4in,keepaspectratio]{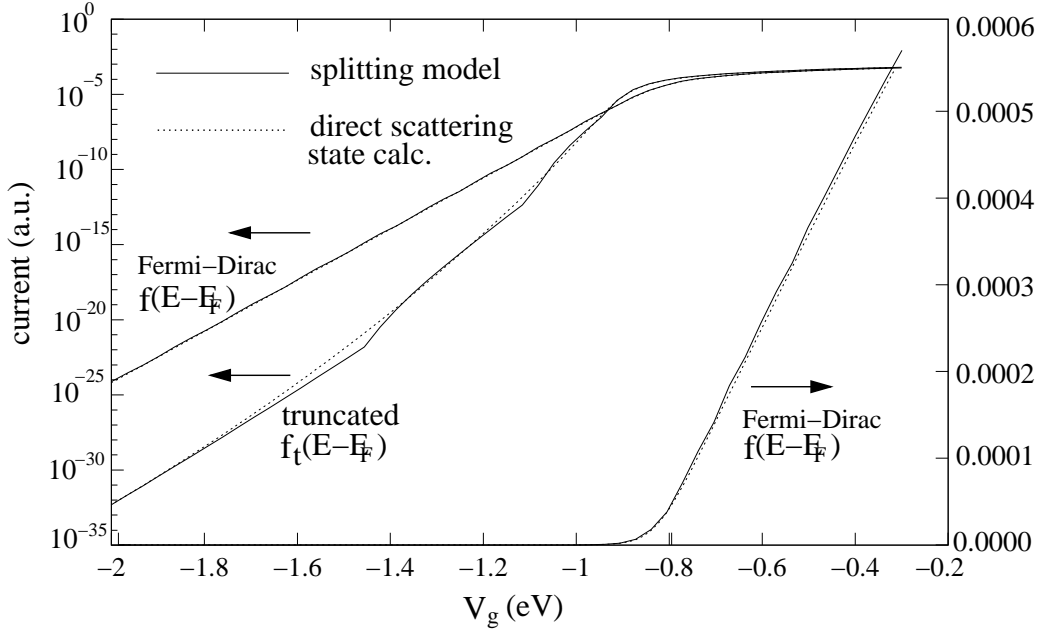}
\caption{ The total current as a function of gate potential $V_g$ as
calculated from equation (14)  in comparison with the current
calculated from equation (4) using scattering states. The $f_t$ is
an artificially sharp truncation for the occupation function to test
the tunneling current.}
\end{figure}

\begin{figure}
\centering
\includegraphics[width=5in,keepaspectratio]{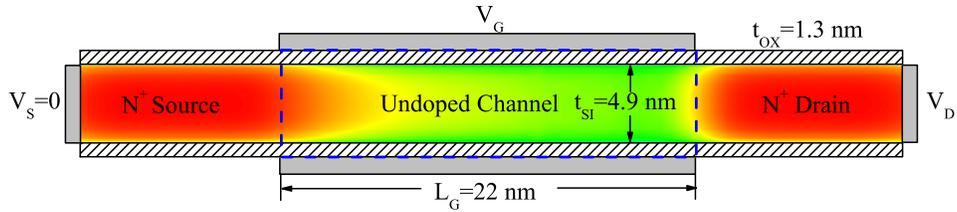}
\caption{(Color online) Structure of the simulated 22 nm Double Gate
Ultra-Thin-Body Filed-Effect Transistor.}
\end{figure}

\begin{figure}
\centering
\includegraphics[width=4in,keepaspectratio]{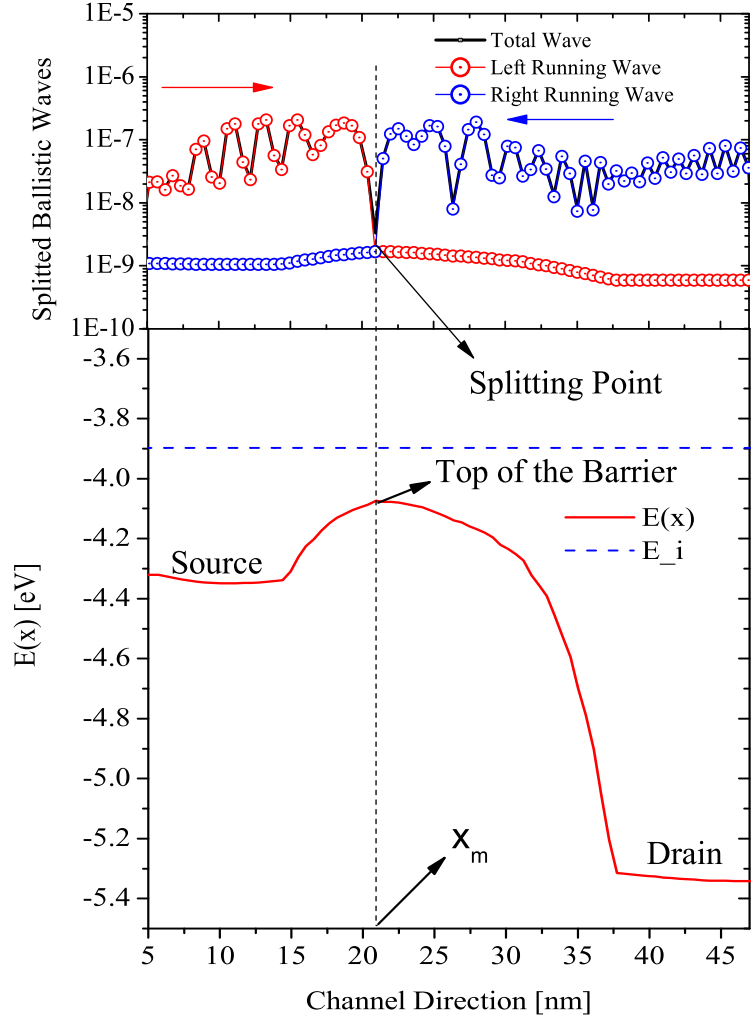}
\caption{(Color online) The splitted wave functions for ballistic
case.}
\end{figure}

\begin{figure}
\centering
\includegraphics[width=4in,keepaspectratio]{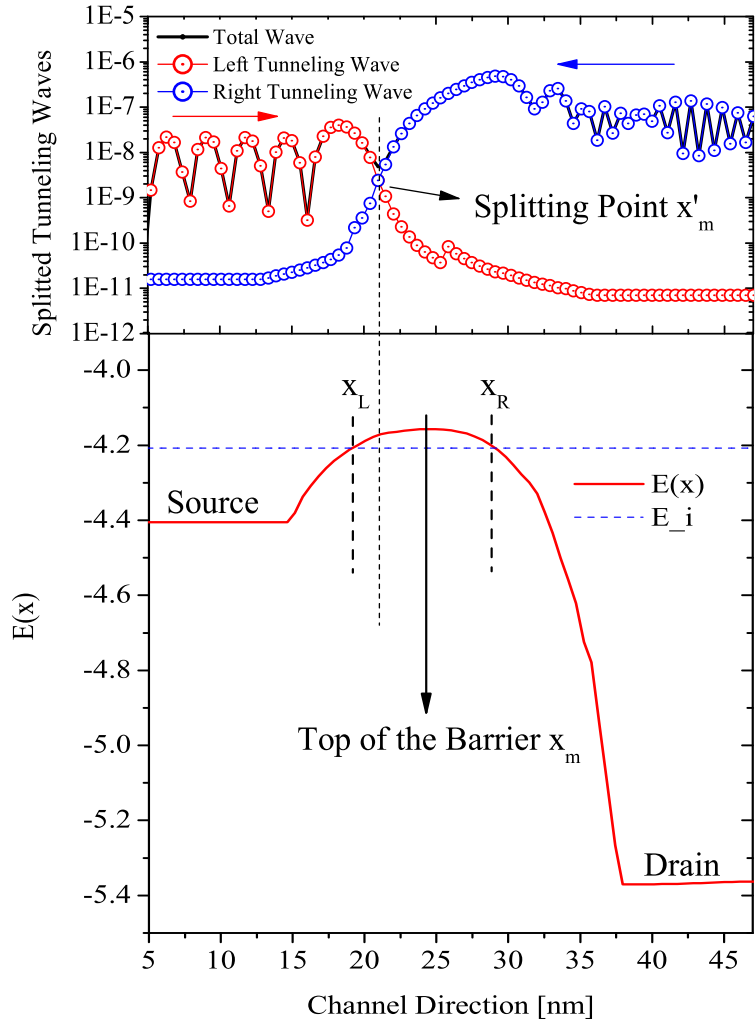}
\caption{(Color online) The splitted wave functions for tunneling
case.}
\end{figure}

\begin{figure}
\centering
\includegraphics[width=4in,keepaspectratio]{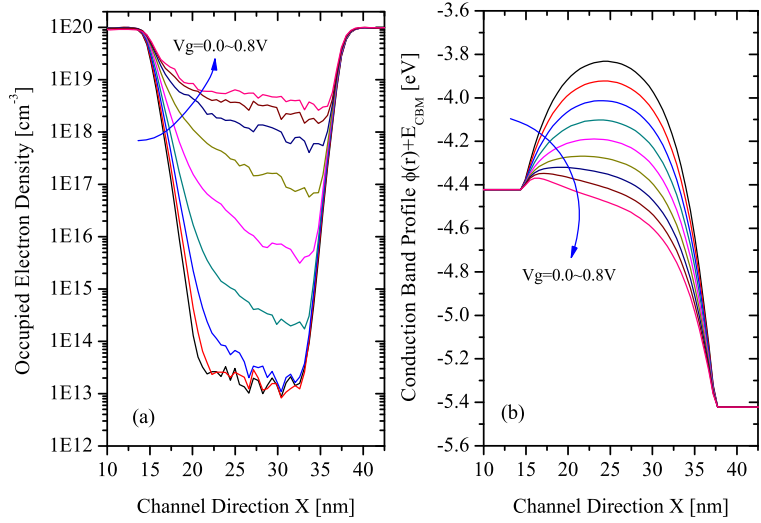}
\caption{(Color online) 1D profile of the occupied electron density
(a) and the conduction band profile (b) for gate bias
$V_g=0.0-0.8V$.}
\end{figure}

\begin{figure}
\centering
\includegraphics[width=4in,keepaspectratio]{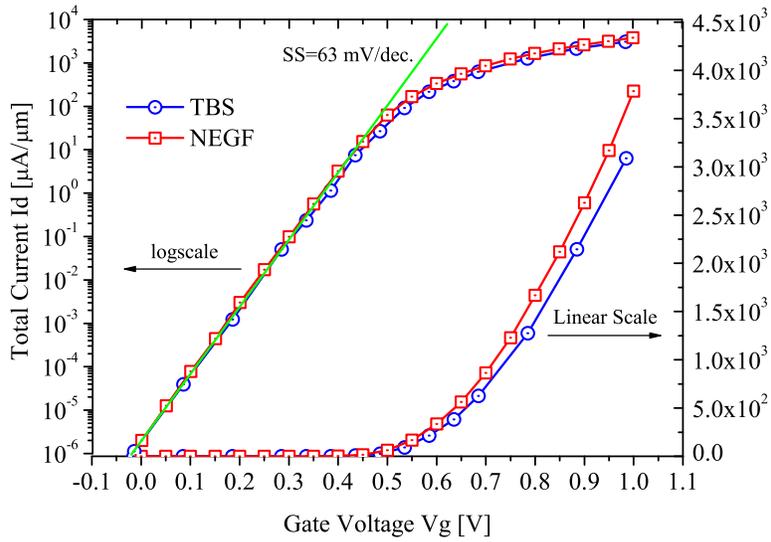}
\caption{(Color online) Transfer characteristics compared with
Ref.41 from the 3D quantum mechanical model.}
\end{figure}

\begin{table}
\begin{center}
\begin{tabular}{|l||c|c|}\hline

  & TBS & TB+NEGF\\ \hline

$I_{ON}$ ($\mu A/\mu m$) & 3265 & 3740\\ \hline

$V_{th}$ (mV) & 460 & 450 \\ \hline

S (mV/dec.) & 63 & 63 \\ \hline
\end{tabular}
\end{center}
\caption{Comparison of some key parameters of device performance
with Ref.41.}
\end{table}

\end{document}